\documentclass[a4paper]{jpconf}
\usepackage{graphicx}
\newcommand{\nmax}{\ensuremath{N_{\rm max}}}
\newcommand{\hw}{\ensuremath{\hbar\omega}}
\begin{document}
\title{Ab initio calculations of $p$-shell nuclei up to N$^2$LO in chiral Effective Field Theory}

\author{Pieter Maris}

\address{Dept. of Physics and Astronomy, Iowa State University, Ames, IA 50011, USA}

\ead{pmaris@iastate.edu}

\begin{abstract}
Nuclear structure and reaction theory are undergoing a major
renaissance with advances in many-body methods, realistic interactions
with greatly improved links to Quantum Chromodynamics, the advent of
high performance computing, and improved computational algorithms.
State-of-the-art two- and three-nucleon interactions obtained from
chiral Effective Field Theory provide a theoretical foundation for
nuclear theory with controlled approximations.  With highly efficient
numerical codes, tuned to the current generation of supercomputers, we
can perform ab-initio nuclear structure calculations for a range of
nuclei to a remarkable level of numerical accuracy, with quantifiable
numerical uncertainties.  Here we present an overview of recent
results for No-Core Configuration Interaction calculations of
$p$-shell nuclei using these chiral interactions up to
next-to-next-to-leading order, including three-body forces.  We show
the dependence of the ground state energies on the chiral order; we
also present excitation spectra for selected nuclei and compare the
results with experimental data.
\end{abstract}

\section{Ab Initio Nuclear Structure and High Performance Computing}

A microscopic theory for the structure and reactions of atomic nuclei
poses formidable challenges for high-performance computing.  A nucleus
with $Z$ protons and $N$ neutrons is a self-bound quantum many-body
system with $A=N+Z$ strongly interacting nucleons.  The interactions
feature both attractive and repulsive contributions along with
significant spin and angular momentum dependence.  Furthermore there
are both short-range and long-range terms in the interaction, and in
addition to nucleon-nucleon (NN) interactions, one also needs suitable
three-nucleon forces (3NFs), and possibly even higher many-body
interactions.  The corresponding Hamiltonian can be written as
\begin{eqnarray}
  {\bf \hat H} &=& \sum_{i<j} \frac{(\vec{p}_i - \vec{p}_j)^2}{2\,m\,A}
  + \sum_{i<j} V_{ij} + \sum_{i<j<k} V_{ijk} + \ldots
\end{eqnarray}
where $m$ is the nucleon mass, which we take to be equal for protons
and neutrons.  The nuclear wave functions are the solutions of the
many-body Schr\"odinger equation
\begin{eqnarray}
  {\bf \hat H} \, \Psi(\vec{r}_1,\ldots,\vec{r}_A) &=& E \, \Psi(\vec{r}_1,\ldots,\vec{r}_A)
\end{eqnarray}
at discrete energy levels $E$.

In No-Core Configuration Interaction (NCCI) nuclear structure
calculations~\cite{Barrett:2013nh} the wave function $\Psi$ of a
nucleus consisting of $A$ nucleons is expanded in an $A$-body basis of
Slater determinants $\Phi_k$ of single-particle wave functions
$\phi_{nljm}(\vec{r})$.  Here, $n$ is the radial quantum number, $l$
the orbital motion, $j$ the total spin from orbital motion coupled to
the intrinsic nucleon spin, and $m$ the spin-projection.  The
Hamiltonian ${\bf \hat H}$ is also expressed in this basis and thus
the many-body Schr\"odinger equation becomes a matrix eigenvalue
problem; for $A > 4$ and NN plus 3N interactions, this matrix is
sparse.  The eigenvalues of this matrix are approximations to the
energy levels, to be compared to the experimental binding energies and
spectra, and the corresponding eigenvectors to the nuclear wave
functions.  Although the wave functions themselves are not observable,
they can be employed to evaluate additional physical observables.

Conventionally, one uses a harmonic oscillator (HO) basis with energy
parameter $\hbar\omega$ for the single-particle wave functions.  A
convenient and efficient truncation of the complete
(infinite-dimensional) basis is a truncation on the total number of HO
quanta: the basis is limited to many-body basis states with $\sum_{A}
N_i \le N_0 + \nmax$, with $N_0$ the minimal number of quanta for that
nucleus and \nmax\ the truncation parameter.  (Even (odd) values of
\nmax\ provide results for natural (unnatural) parity.)  Numerical
convergence toward the exact results for a given Hamiltonian is
obtained with increasing \nmax, and is marked by approximate
\nmax\ and \hw\ independence.  In practice we use extrapolations to
estimate the binding energy in the complete (but infinite-dimensional)
space~\cite{Maris:2008ax,Coon:2012ab,Furnstahl:2012qg,More:2013rma,Wendt:2015nba},
based on a series of calculations in finite bases.

The rate of convergence depends both on the nucleus and on the
interaction.  For realistic interactions, the dimension of the matrix
needed to reach a sufficient level of convergence is in the billions,
and the number of nonzero matrix elements is in the tens of trillions,
which saturates available storage on current computing facilities.
All NCCI calculations presented here were performed on the Cray XC30
Edison and Cray XC40 Cori at NERSC and the IBM BG/Q Mira at Argonne
National Laboratory, using the code
MFDn~\cite{doi:10.1002/cpe.3129,SHAO20181}.

\section{Nuclear Interactions from Chiral Effective Field Theory}

Chiral Effective Field Theory ($\chi$EFT) allows us to derive nuclear
interactions (and the corresponding electroweak current operators) in
a systematic
way~\cite{Weinberg:1990rz,Epelbaum:2008ga,Machleidt:2011zz}.  The
chiral expansion is by no means unique: e.g.  different choices for
the functional form of the regulator and/or different choices for the
degrees of freedom lead to different $\chi$EFT interactions.  With the
LENPIC
collaboration~\cite{Binder:2015mbz,Binder:2018pgl,Epelbaum:2018ogq} we
use the same $\chi$EFT interactions for ab initio calculations ranging
from nucleon-nucleon and nucleon-deuteron scattering to the structure
of medium-mass nuclei.  Specifically, here we use the semilocal
coordinate-space regularized chiral potentials of
Refs.~\cite{Epelbaum:2014efa,Epelbaum:2014sza} to calculate the
binding energies and spectra of $p$-shell nuclei.  The leading order
(LO) and next-to-leading order (NLO) contributions are given by
NN-only potentials while 3NFs appear first at next-to-next-to-leading
order (N$^2$LO) in the chiral
expansion~\cite{Epelbaum:2008ga,Machleidt:2011zz}.  Four-nucleon
forces are even more suppressed and start contributing at N$^3$LO.
The chiral power counting thus provides a natural explanation of the
observed hierarchy of nuclear forces.

The Low-Energy Constants (LECs) in the NN-only potentials of
Refs.~\cite{Epelbaum:2014efa,Epelbaum:2014sza} have been fitted to
nucleon-nucleon scattering, without any input from nuclei with $A>2$.
The 3NFs at N$^2$LO involve two LECs which govern the strength of the
one-pion-exchange-contact term and purely contact 3NF contributions.
Conventionally, these LECs are expressed in terms of two dimensionless
parameters $c_D$ and $c_E$.  Obviously, these LECs cannot be fixed
from nucleon-nucleon scattering; they have to be fitted to select
3-body (or higher $A$-body) observables.  We follow the commonly
adopted
practice~\cite{Epelbaum:2002vt,Nogga:2005hp,Navratil:2007we,Gazit:2008ma}
and use the $^3$H binding energy as one of the observables; this gives
us a correlation between $c_D$ and $c_E$.

A wide range of observables has been considered in the literature to
constrain the remaining LEC.  In Ref.~\cite{Epelbaum:2018ogq}
different ways to fix this LEC in the 3-nucleon sector were explored,
and it was shown that it can be reliably determined from the minimum
in the differential cross section in elastic nucleon-deuteron
scattering at intermediate energies.  This allows us to make
parameter-free calculations for $A \ge 4$ nuclei.  In these
proceedings we present an overview of the ground state energies for
all stable $p$-shell nuclei (excluding mirror nuclei), as well as
excitation spectra for selected nuclei up to $A=12$, all obtained with
the same semilocal regulator $R=1.0$~fm and the same LECs.
Specifically, the LECs values for the 3NFs at N$^2$LO are $c_D=7.2$
and $c_E=-0.671$, as determined in Ref.~\cite{Epelbaum:2018ogq}.
Application of these interactions to nucleon-deuteron scattering can
be found in Refs.~\cite{Binder:2015mbz,Binder:2018pgl} for NN-only
potentials, along with selected properties of light- and medium-mass
nuclei, and in Ref.~\cite{Epelbaum:2018ogq} including the 3NFs at
N$^2$LO.

\section{Ground State Energies for $p$-shell Nuclei}

\begin{figure}[b]
  \center\includegraphics[width=0.9\columnwidth]{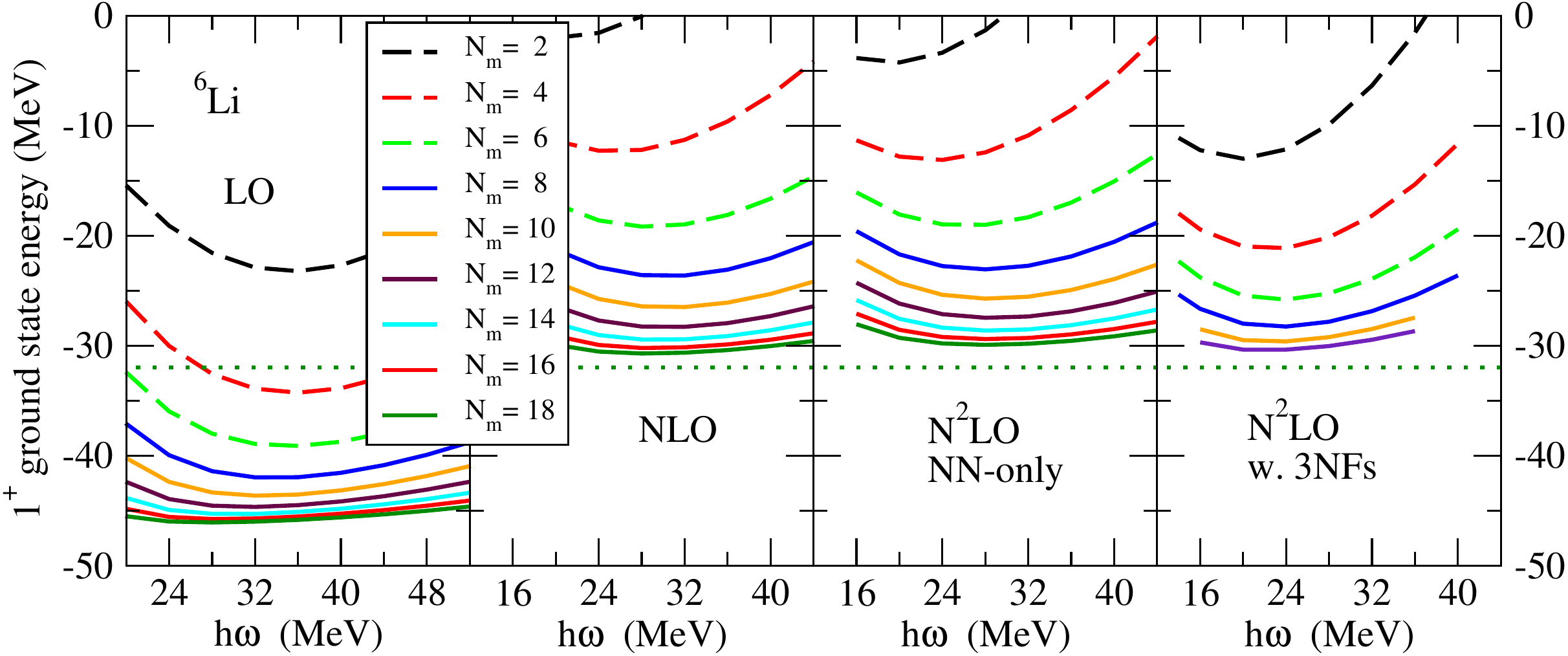}
  \caption{\label{Fig:res_gs_6Li_hw}
    (Color online) Calculated ground state energy of $^6$Li using chiral
    LO, NLO, and N$^2$LO interactions at $R=1.0$~fm as function of the
    basis HO parameter \hw\ for $\nmax=2$ to $18$ for NN-only potentials
    (left 3 panels) and at N$^2$LO w. 3NFs, SRG-evolved to
    $\alpha=0.02$~fm$^4$ for $\nmax=2$ to $12$ (right-most panel).
    The dotted horizontal line is the experimental value.}
\end{figure} 
Here we present our results for the ground state energies of the
stable $p$-shell nuclei, excluding mirror nuclei, all obtained with
the same semilocal chiral interactions up to N$^2$LO.  In
Fig.~\ref{Fig:res_gs_6Li_hw} we show the ground state ($J^P=1^+$)
energy of $^6$Li as function of the HO basis parameter \hw\ for a
range of \nmax\ values.  With NN-only potentials, we can perform
calculations up to $\nmax=18$ for $A=6$ nuclei.  This is sufficient to
achieve a reasonable level of convergence, as can be seen from the
left three panels of Fig.~\ref{Fig:res_gs_6Li_hw}.  With 3NFs however,
we are limited to significantly smaller bases, and in order to improve
the numerical convergence with basis size we therefore first perform a
Similarity Renormalization Group (SRG)
transformation~\cite{Bogner:2007rx,Bogner:2009bt,Roth:2013fqa} on the Hamiltonian.
The right-most panel of Fig.~\ref{Fig:res_gs_6Li_hw} shows results for
the ground state energy of $^6$Li at N$^2$LO including 3NFs at a very
modest SRG flow parameter $\alpha=0.02$~fm$^4$ (note that $\alpha=0$
correspond to the original Hamiltonian, without SRG), for calculations
up to $\nmax=12$.  Indeed, the convergence with increasing \nmax\ is
significantly improved with this SRG-evolved interaction compared to
the bare NN-only interactions at NLO and N$^2$LO.  At $\nmax=12$ the
level of convergence is already comparable to that of the bare NLO and
N$^2$LO potentials at $\nmax=16$.  Also note that the variational
minimum in \hw\ shifts to lower values due to the SRG evolution.

\begin{figure}[tbh]
  \includegraphics[width=0.46\columnwidth]{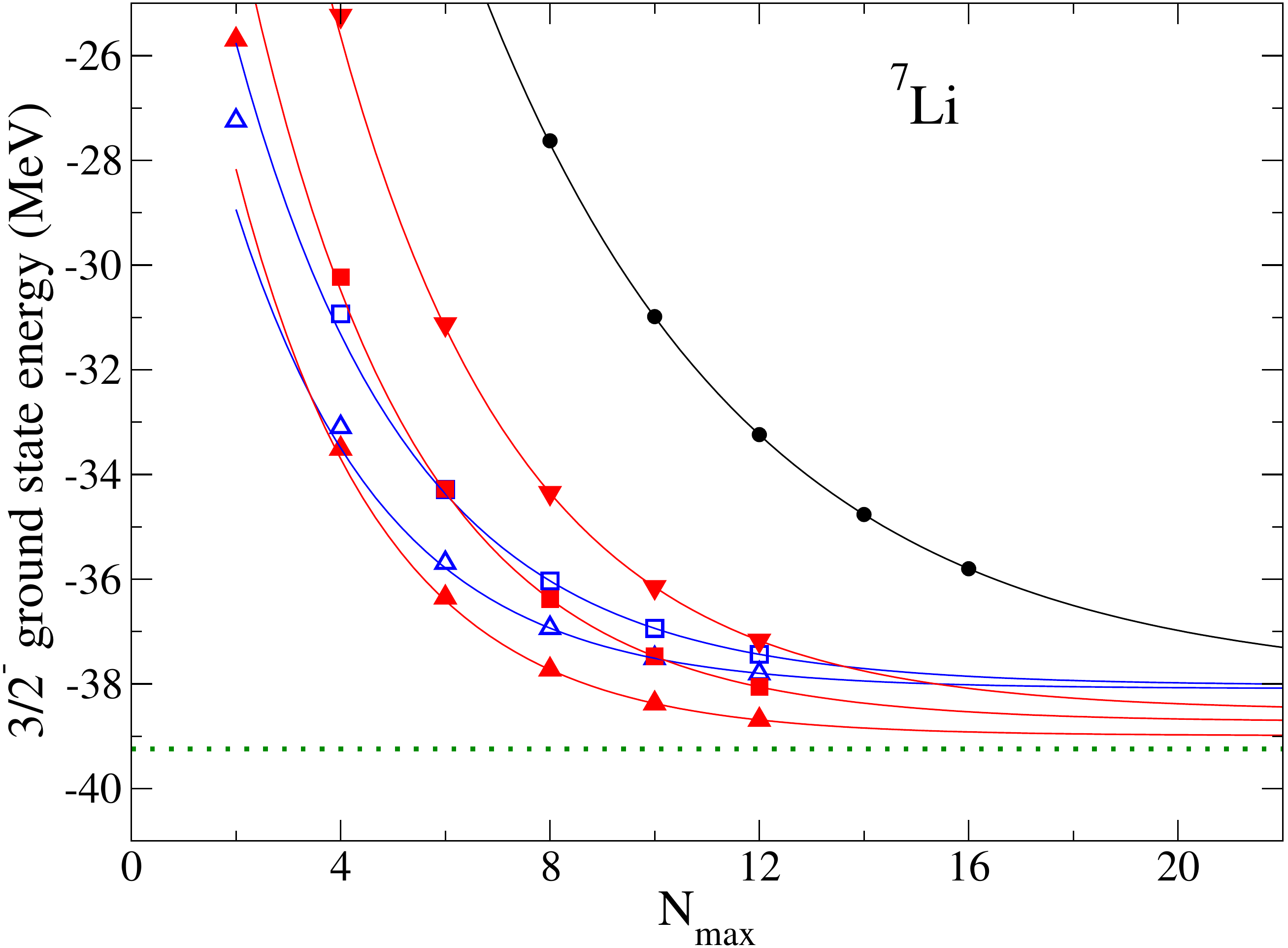}
  \qquad
  \includegraphics[width=0.46\columnwidth]{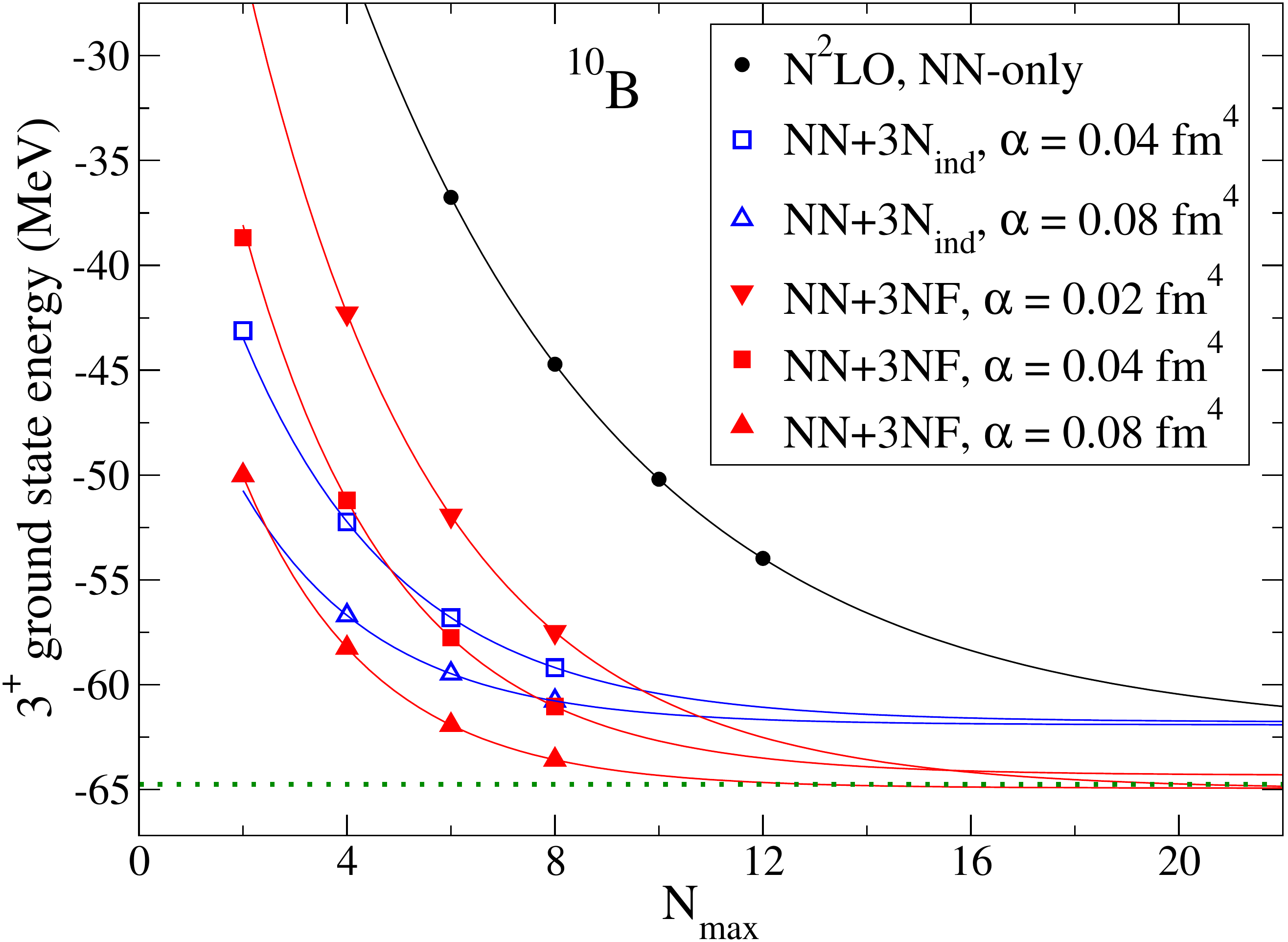}
  \caption{\label{Fig:res_gs_7Li10B_Nm}
    (Color online) Calculated ground state energies using chiral
    N$^2$LO interactions at $R=1.0$~fm as function of \nmax\ at the
    variational minimum in \hw\ for $^7$Li (left) and $^{10}$B (right).
    The dotted horizontal line is the experimental value.}
\end{figure} 
In Fig.~\ref{Fig:res_gs_7Li10B_Nm} we show the ground state energies
of $^7$Li (left, $J^P=\frac{3}{2}^-$) and $^{10}$B (right, $J^P=3^+$)
as function of \nmax\ at fixed \hw\ values close to the variational
minimum with the N$^2$LO interaction with and without explicit 3NFs.
Based on these results in finite bases, we can use extrapolations to
the complete (infinite-dimensional) basis.  Here we use a three
parameter fit at fixed \hw\ at or just above the variational minimum
\begin{eqnarray}
   E(N_{\max}) &\approx&  E_{\infty} + a \exp{(-b N_{\max})} \,,
\end{eqnarray}
which seems to work well for a range of interactions and
nuclei~\cite{Maris:2008ax,Maris:2013poa,Jurgenson:2013yya}.  The lines
in Fig.~\ref{Fig:res_gs_7Li10B_Nm} correspond to the extrapolating
function fitted to the three highest available \nmax\ values.

Again, with the SRG-evolved interactions the ground state energies
converge more rapidly with \nmax\ than with the bare (black dots and
curves) NN-only N$^2$LO interaction.  However, as a consequence of the
SRG transformation, our results do depend on the SRG flow parameter
$\alpha$, because we do not incorporate any induced interactions
beyond 3NFs.  Without explicit 3NFs, this dependence seems to be
negligible, and typically less than the extrapolation uncertainty --
the bare NN-only N$^2$LO interaction and the two SRG-evolved
interaction with induced 3NFs extrapolate to approximately the same
value.  On the other hand, with explicit 3NFs there is a weak but
noticeable dependence on the SRG parameter $\alpha$, as can be seen by
the spread of the red extrapolation curves in
Fig.~\ref{Fig:res_gs_7Li10B_Nm}.  This $\alpha$ dependence is due to
induced 4-body (and higher-body) interactions which we have neglected.

\begin{table}[t]
  \caption{\label{Tab:res_gs_Eb} Ground state energies of stable $A =
    4$ to $16$ nuclei with $\chi$EFT interactions up to N$^2$LO using
    $R=1.0$~fm~\cite{Binder:2018pgl,Epelbaum:2018ogq}.
    The uncertainty estimate is only the extrapolation uncertainty in
    the many-body calculation, and does not include the chiral
    truncation error, nor uncertainties in the LECs.  Entries with an
    asterix $^*$ indicate excited states for nuclei where the
    calculated and experimental ground states have different $J^P$.
  Experimental values are extracted from Ref.~\cite{Audi2003337}.}
  \lineup
  \begin{center}
    \begin{tabular}{ll|lllll|l}
      \br
      \multicolumn{2}{l|}{Nucleus} & LO & NLO & N$^2$LO &  \multicolumn{2}{c|}{N$^2$LO including 3NFs} &
      \\
               & $J^P$ & NN-only & NN-only & NN-only & $\alpha=0.04$~fm$^4$ & $\alpha=0.08$~fm$^4$ & expt.
      \\
      \mr
      $^4$He   & $0^+$ & $-45.453(6)$ & $-28.533(4)$ & $-28.11(1)$ & $-28.202(5)$ & $-28.298(2)$ & $-28.296$
      \\
      $^6$He   & $0^+$ & $-43.2(2)\0\0$&$-28.7(2)\0\0$&$-27.9(2)\0$&$-28.55(15)$&$-28.79(8)$ &$-29.27\0$
      \\
      $^6$Li   & $1^+$ & $-46.7(1)$    &$-31.6(2)$    & $-31.0(2) $&$-31.49(16)$&$-31.72(6)$ & $-31.99$
      \\
      $^7$Li   & $\frac{3}{2}^-$ & $-57.1(2)^*$& $-38.7(3)$ & $-38.0(4)$&$-38.72(16)$&$-38.99(6)$& $-39.24$
      \\
      $^8$He   & $0^+$ & $-39.8(6)$    & $-29.7(5)$   & $-27.8(6)$ & $-29.5(3)$ & $-29.9(2)$ & $-31.41$
      \\
      $^8$Li   & $2^+$ & $-55.7(5)$    & $-40.3(7)$   & $-39.0(8)$ & $-40.4(4)$ & $-40.7(2)$ & $-41.28$
      \\
      $^8$Be   & $0^+$ & $-87.7(4)$    & $-56.0(7)$   & $-55.4(9)$ & $-55.6(5)$ & $-56.1(3)$ & $-56.50$
      \\
      $^9$Li   & $\frac{3}{2}^-$ & $-57.1(4)$ & $-43.9(7)$ & $-41.7(8)$ &$-43.9(4)$&$-44.0(2)$& $-45.34$
      \\
      $^9$Be   & $\frac{3}{2}^-$ & $-84.7(7)$ & $-58.0(1.4)$&$-56.4(1.5)$&$-57.5(5)$&$-58.0(3)$& $-58.16$
      \\
      $^{10}$Be & $0^+$ & $-92.2(8)$ & $-65.2(1.5)$ & $-62.8(1.7)$ & $-64.1(9)$ & $-64.9(5)$ & $-64.98$ 
      \\
      $^{10}$B  & $3^+$ & $-88.1(1.2)^*$&$-64.6(1.5)^*$& $-62.3(1.7)^*$&$-64.3(8)$ & $-64.9(5)$ & $-64.75$
      \\
      $^{10}$B  & $1^+$ & $-93.9(8)$   & $-64.9(1.8)$ & $-63.1(1.9)$ &$-63.1(1.0)^*$&$-64.1(8)^*$& $-64.03^*$
      \\
      \mr
      & & \multicolumn{3}{c|}{SRG evolved to $\alpha=0.04$~fm$^4$ } & $\alpha=0.04$~fm$^4$ & $\alpha=0.08$~fm$^4$ &
      \\      
      \mr
      $^{11}$Be & $\frac{1}{2}^+$ & --- & --- & --- & $-64.7(1.3)^*$ & $-65.4(8)^*$ & $-65.48$
      \\
      $^{11}$Be & $\frac{1}{2}^-$ & --- & --- & --- & $-65.8(1.2)$   & $-65.7(8)$   & $-65.16^*$
      \\
      $^{11}$B  & $\frac{3}{2}^-$ & $-108.(1.)$ & $-76.8(6)$ & $-73.9(7)$ & $-77.2(9)$ & $-77.7(5)$ & $-76.21$ 
      \\
      $^{12}$Be & $0^+$ & --- & --- & --- &  $-68.9(1.4)$ & $-69.8(9)$ & $-68.65$
      \\
      $^{12}$B  & $1^+$ & $-111.(1.)^*$ & $-82.6(8)$   & $-78.6(8)$  & $-81.9(9)^*$ & $-82.5(5)^*$ & $-79.58$
      \\
      $^{12}$B  & $2^+$&  $-111.(1.)^*$ & $-82.3(9)^*$ & $-77.8(7)^*$ & $-82.8(9)$   & $-83.2(5)$  & $-78.63^*$
      \\
      $^{12}$C  & $0^+$ & $-139.(1.)$ & $-95.5(7)$ & $-92.7(6)$ & $-94.7(1.0)$ & $-95.5(5)$   & $-92.16$ 
      \\
      $^{13}$B  &$\frac{3}{2}^-$& --- & --- & --- & $-89.5(1.0)$  & $-90.3(7)$  & $-84.45$ 
      \\
      $^{13}$C  &$\frac{1}{2}^-$& --- & --- & --- & $-104.7(1.0)$ & $-104.4(4)$ & $-97.11$
      \\
      $^{14}$C  & $0^+$ & --- & --- & --- & $-116.0(1.3)$ & $-116.1(5)$ & $-105.28$ 
      \\
      $^{14}$N  & $1^+$ & --- & --- & --- & $-117.3(1.3)$ & $-117.4(4)$ & $-104.66$ 
      \\
      $^{15}$N  &$\frac{1}{2}^-$& --- & --- & --- & $-130.4(1.6)$ & $-131.0(6)$ & $-115.49$ 
      \\
      $^{16}$O  & $0^+$ & $-223.2(4)$ & $-152.(1.)$ & $-146.(1.)$ & $-144.(2.)$ & $-145.2(8)$ & $-127.62$
      \\
      \br
  \end{tabular}
  \end{center}
\end{table}
In Table~\ref{Tab:res_gs_Eb} we summarize our results up to N$^2$LO
for the ground state energies of stable $p$-shell nuclei, excluding
mirror nuclei, extrapolated to the complete basis.  Our estimate of
the extrapolation uncertainty is based on the difference with smaller
\nmax\ extrapolations, as well as the basis \hw\ dependence over an
$8$ to $12$~MeV span in \hw\ values around the variational minimum,
adjusted to be at least 20\% of the difference with the variational
minimum~\cite{Binder:2018pgl}.

With NN-only potentials we use the bare interaction up to $A=10$, for
which we can perform calculation at $\nmax=12$ or higher.  For select
nuclei with $11 \le A \le 16$ we use the SRG-evolved interaction at
$\alpha=0.04$~fm$^4$ with induced 3NFs for NN-only potentials up to
$\nmax=8$.  At N$^2$LO with explicit 3NFs we present results with
SRG-evolved interactions at both $\alpha=0.04$~fm$^4$ and
$\alpha=0.08$~fm$^4$.  As expected, the calculations at
$\alpha=0.08$~fm$^4$ are betted converged, and have therefore a
smaller extrapolation uncertainty than those at $\alpha=0.04$~fm$^4$.
The anticipated $\alpha$ dependence appears to be of the same order
of magnitude as the extrapolation uncertainty.

Generally, the agreement with the experimental binding energies
improves as one goes from LO to NLO to N$^2$LO.  At LO all $p$-shell
nuclei are significantly overbound, but at N$^2$LO the binding
energies of nuclei up to $A=12$ are within few percent of the
experimental values.  As $A$ increases beyond $A=12$, the nuclei
become more and more overbound -- $^{12}$C is overbound by about 3\%
whereas $^{16}$O is overbound by about 13\%.  The overbinding of
$^{16}$O is significantly larger than the estimated chiral truncation
uncertainty~\cite{Binder:2018pgl}, even with the inclusion of the
explicit 3NFs~\cite{Epelbaum:2018ogq}, and it is as of yet unclear
what the origin of this overbinding is.

At NLO and higher, we obtain the correct spin and parity for the
ground states of most $p$-shell nuclei -- the exceptions are $^{10}$B,
$^{11}$Be, and $^{12}$B, for which we include both the experimental
and the calculate ground states in Table~\ref{Tab:res_gs_Eb}.  For
$^{10}$B, the NN-only interactions produce a $J^P=1^+$ ground state,
whereas the experimental ground state has $J^P=3^+$.  With the
consistent explicit 3NFs at N$^2$LO we are able to reproduce the
experimental ground state for $^{10}$B, in agreement with previous
studies of $^{10}$B with $\chi$EFT
interactions~\cite{Navratil:2007we,Jurgenson:2013yya}.  For $^{12}$B
the situation is the opposite: at NLO and N$^2$LO without the 3NFs we
do find the correct ground state, $J^P=1^+$, but adding the 3NFs to
the N$^2$LO NN potential leads to a ground state with $J^P=2^+$, and
the $J^P=1^+$ state becomes the first excited state, with an
excitation energy of about 1 MeV.  It remains to be seen whether or
not this discrepancy gets resolved at higher order in the chiral
expansion.

The situation in $^{11}$Be is different: here we have a nucleus with
{\em parity inversion}, that is, the ground state has the opposite
parity of what one would expect based on the shell-model.  In NCCI
calculations the 'natural' and 'unnatural' parity states are expressed
in bases with even or odd \nmax\ values respectively.  For $^{11}$Be
that means the negative parity states are calculated in bases with
even \nmax\ and the positive parity states states in bases with odd
\nmax.  We then perform an extrapolation to the complete basis for the
lowest state with even \nmax\ as well as for the lowest state with odd
\nmax.  This leads to the energies listed in Table~\ref{Tab:res_gs_Eb}
for the $\frac{1}{2}^+$ state (the experimental ground state) and for
the the $\frac{1}{2}^-$ state (the lowest natural parity state).
Although the latter has a lower energy in our calculations, the
difference with that of the $\frac{1}{2}^+$ is less than the
extrapolation uncertainty, and within their uncertainties, both
energies agree with the experimental values.  In order to reliably
determine which of these two states is the ground state we should use
more sophisticated calculational methods for this system and follow
e.g. the approach discussed in Ref.~\cite{Calci:2016dfb} for $^{11}$Be.

\section{Excitation Spectra for $p$-shell Nuclei}

\begin{figure}[b]
  \center\includegraphics[width=0.9\columnwidth]{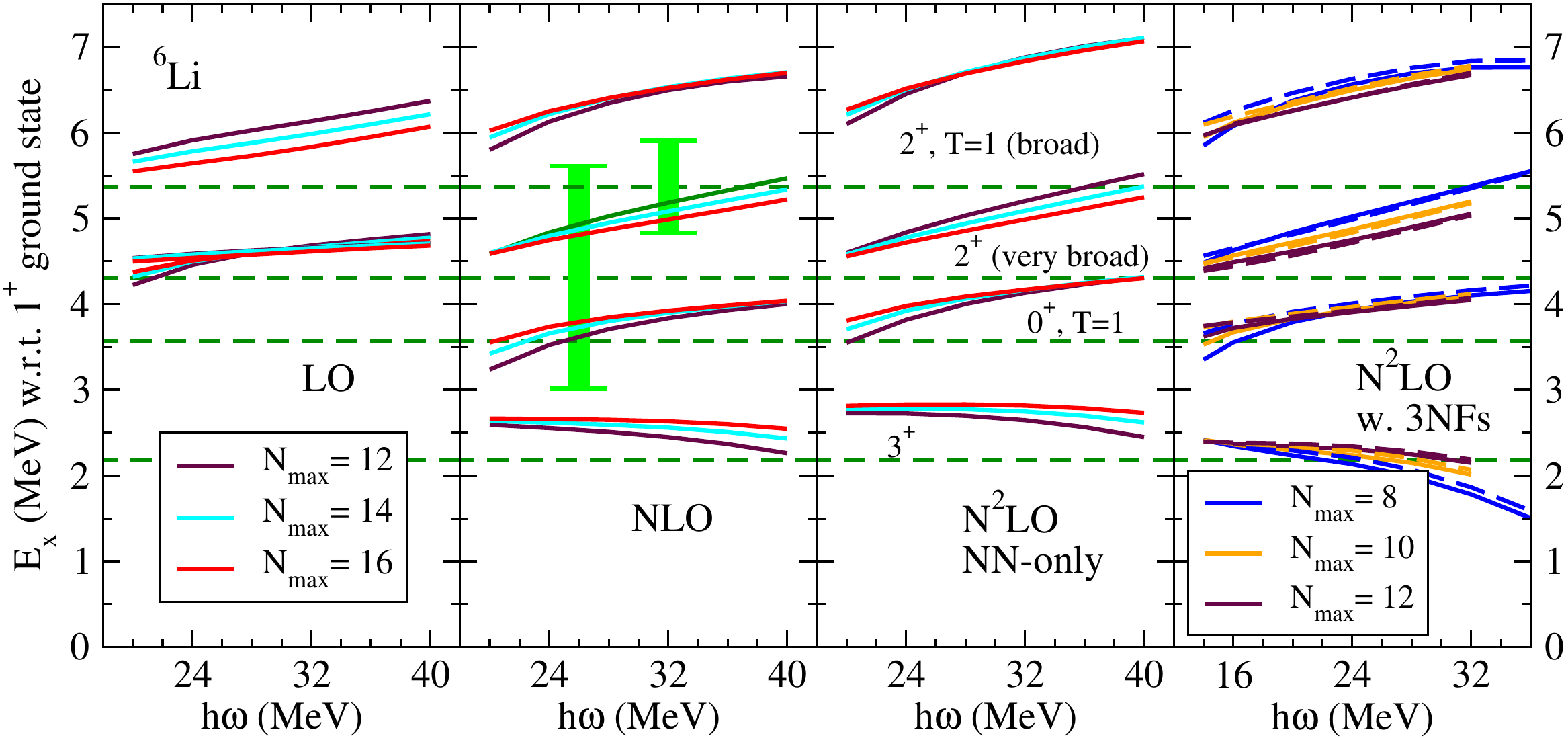}
  \center\includegraphics[width=0.9\columnwidth]{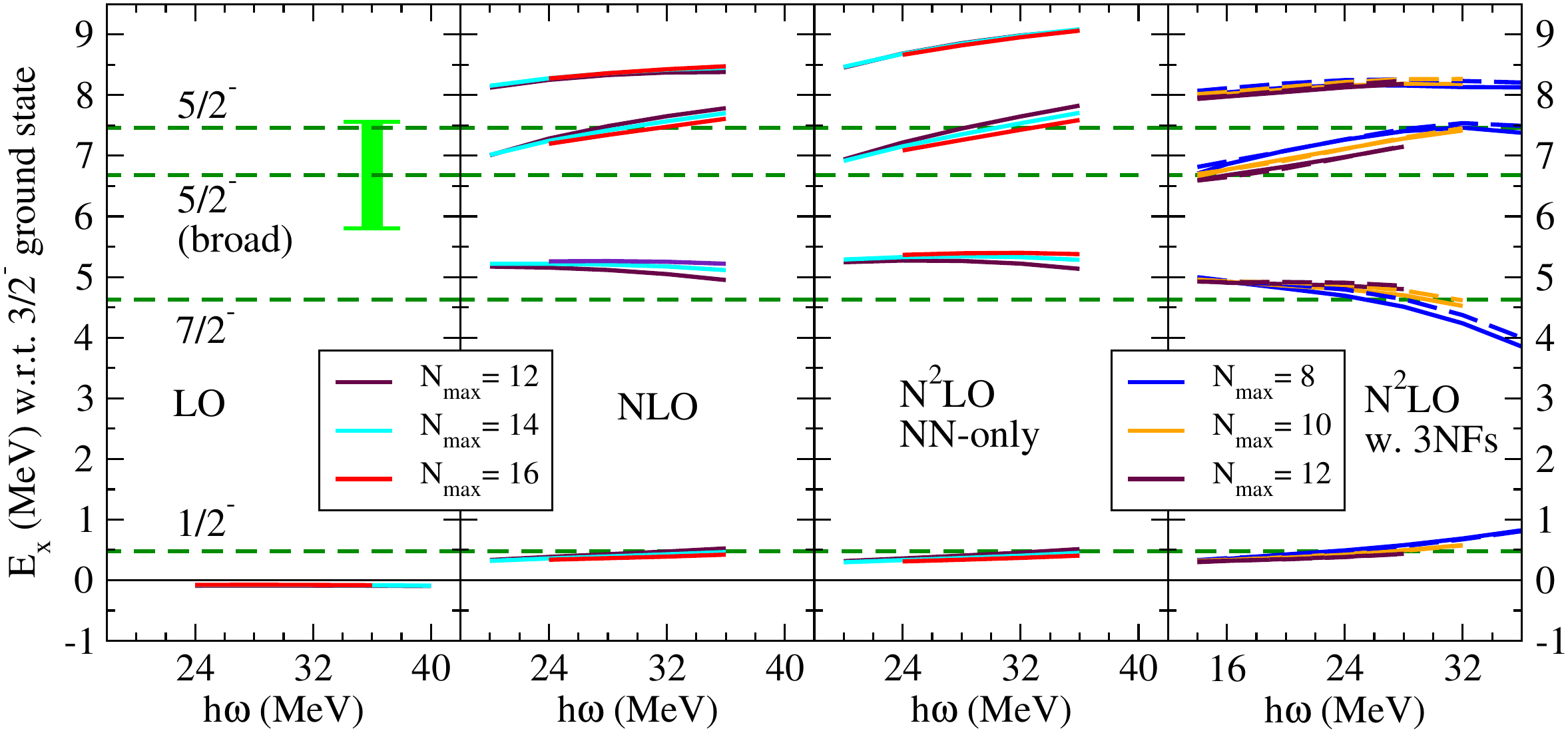}
  \caption{\label{Fig:res_spectrum_6Li7Li_hw}
    (Color online) Calculated excitation spectrum of $^6$Li (top) and $^7$Li (bottom)
    using chiral LO, NLO, and N$^2$LO interactions at $R=1.0$~fm as
    function of the basis HO parameter \hw\ for $\nmax=12$ to $16$ for
    NN-only potentials (left 3 panels) and at N$^2$LO w. 3NFs,
    SRG-evolved to $\alpha=0.04$~fm$^4$ (solid) and
    $\alpha=0.08$~fm$^4$ (dashed) for $\nmax=8$ to $12$ (right-most
    panels).  The dashed horizontal lines are the experimental values~\cite{TILLEY20023}.}
\end{figure}
In addition to the ground state energies, we also obtain the energy
levels of excited states.  The energy differences with the ground
state generally converge significantly better than the actually
binding energies of excited states, at least for states of the same
parity.  In Fig.~\ref{Fig:res_spectrum_6Li7Li_hw} we show the
low-lying spectra of $^6$Li and $^7$Li as function of the HO basis
parameter \hw\ for several of \nmax\ values.  Again, with NN-only
potentials we achieve a reasonable level of convergence, in particular
for narrow excited states like the $3^+$ state in $^6$Li and the
$\frac{1}{2}^-$ and $\frac{7}{2}^-$ states in $^7$Li.  The persistent
increase of the excitation energies of with increasing \hw\ for the
higher excited states suggest that these are (significantly) broader,
and therefore poorly converging in a HO basis.  Indeed, the two $2^+$
states in $^6$Li are broad; and although the $0^+$ in $^6$Li (the
analog state of $^6$He) is narrow, in our calculations with NN-only
interactions up to N$^2$LO, $^6$He is not or barely bound, see
Table~\ref{Tab:res_gs_Eb}; hence, with these interactions this state
will be broad and poorly converging.

At LO the spectra do not agree with experiment -- most excitation
energies are too large, and often the order of the states is
incorrect: e.g. in $^7$Li the ground state, $\frac{3}{2}^-$, and the
first excited state, $\frac{1}{2}^-$, are essentially degenerate.
Indeed, the LO potential is not very realistic -- not only is it
significantly too attractive (it overbinds all $p$-shell nuclei by up
to a factor of two), it is also missing e.g. essential spin-orbit
couplings.  However, starting at NLO the spectra tend to be in
qualitative agreement with data.
At N$^2$LO with explicit 3NFs we use SRG evolution to improve
convergence of the NCCI calculations.  The dependence of the
excitation energies on the SRG parameter $\alpha$ is negligible, much
smaller than the \hw\ dependence, as can be seen in the the right-most
panels of Fig.~\ref{Fig:res_spectrum_6Li7Li_hw}.  Generally, inclusion
of the 3NFs improves agreement with experiment (see also Fig.~9 of
Ref.~\cite{Epelbaum:2018ogq}).  In particular, we see in
Fig.~\ref{Fig:res_spectrum_6Li7Li_hw} that the excitation energy of
the $3^+$ state of $^6$Li moves slightly closer to experiment; and in
$^7$Li the $\frac{7}{2}^-$ also moves slightly closer to experiment.
Furthermore the second $\frac{5}{2}^-$ state becomes much better
converged while the first $\frac{5}{2}^-$ exhibits a persistent
\hw\ dependence, suggesting that the first $\frac{5}{2}^-$ is broad,
and the second narrow, both in agreement with data.

\begin{figure}[b]
  \includegraphics[width=0.3\columnwidth]{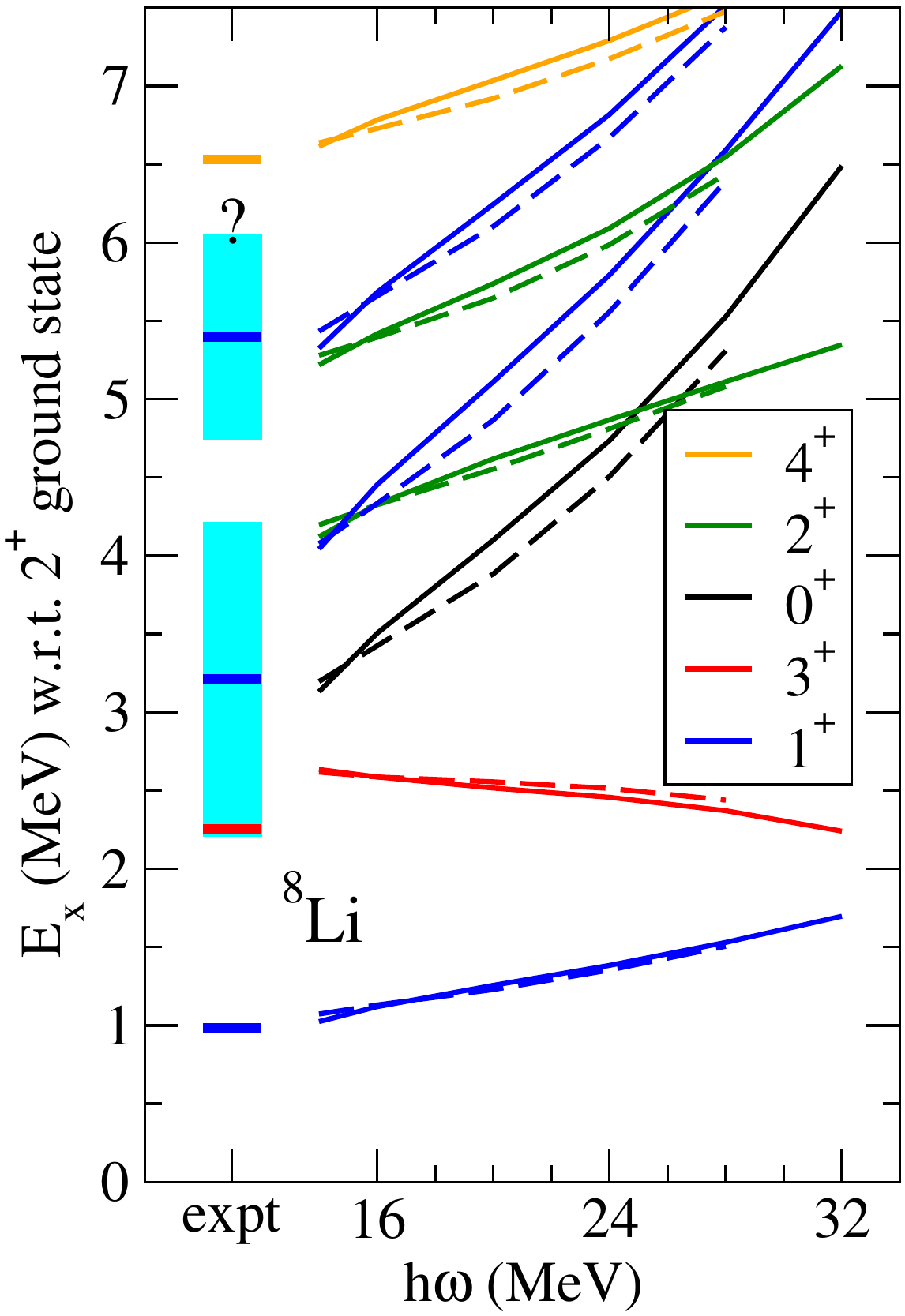}
  \quad
  \includegraphics[width=0.31\columnwidth]{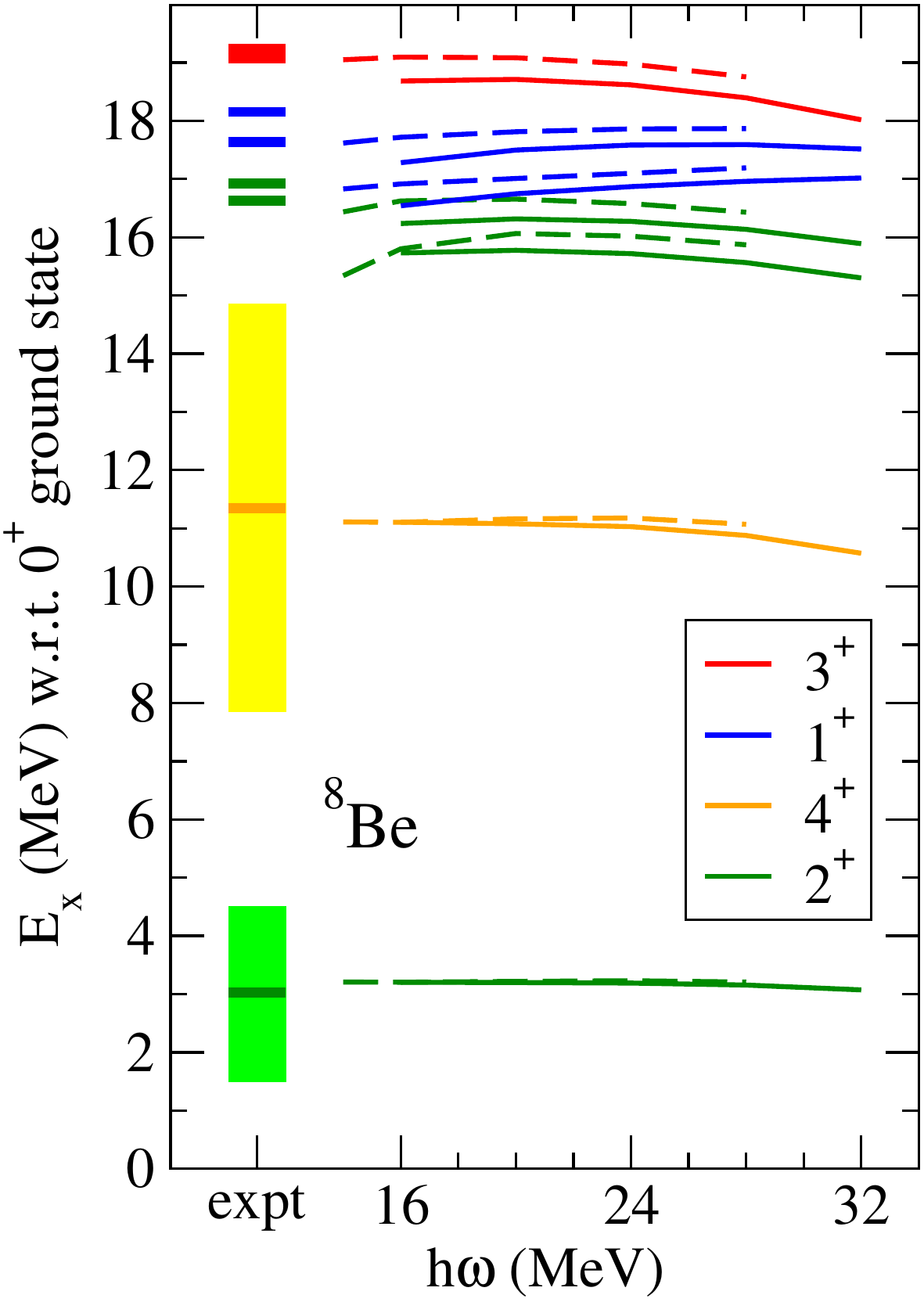}
  \quad
  \includegraphics[width=0.3\columnwidth]{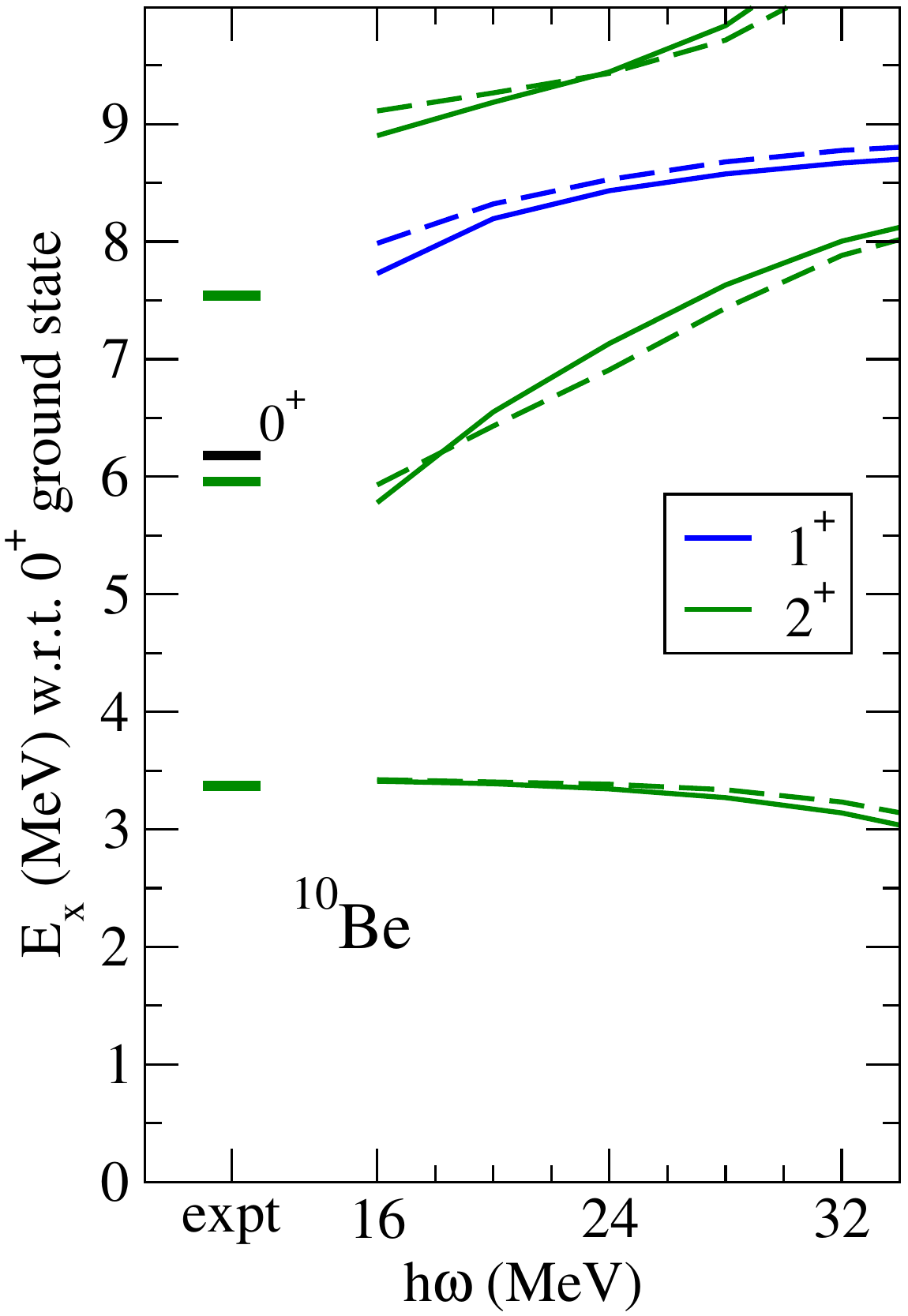}
  \caption{\label{Fig:res_spectra_hw}
    (Color online) Positive-parity excitation spectra of $^8$Li,
    $^8$Be, and $^{10}$Be using the chiral N$^2$LO interaction
    w. 3NFs, SRG-evolved to $\alpha=0.04$~fm$^4$ (solid) and
    $\alpha=0.08$~fm$^4$ (dashed), as function of the
    basis HO parameter \hw, with $\nmax=10$ for $^8$Li
    and $^8$Be and $\nmax=8$ for $^{10}$Be.
    Experimental levels from ENSDF, Ref.~\cite{TILLEY2004155}.}
\end{figure} 
In Fig.~\ref{Fig:res_spectra_hw} we show the low-lying positive-parity
spectra for $^8$Li, $^8$Be, and $^{10}$Be at N$^2$LO with explicit
3NFs, SRG evolved to $\alpha = 0.04$~fm$^4$ (solid) and $0.08$~fm$^4$
(dashed).  Again, the SRG dependence is negligible compared to the
\hw\ dependence, except for the high-lying pairs of $2^+$, $1^+$, and
$3^+$ states in $^8$Be; given this SRG dependence, the spectrum of
$^8$Be is in quite reasonable agreement with the data.  For $^8$Li we
do find the known narrow $1^+$, $3^+$, and $4^+$ states, as well as
two poorly converged (i.e. broad) $1^+$ states, all in reasanoble
agreement with experiment; in addition we find one $0^+$ state, as
well as two $2^+$ states, all poorly converged.

The first excited state in $^{10}$Be, with $J^P=2^+$, is quite well
converged, and in excellent agreement with the experimental excitation
energy.  We also do find two additional $2^+$ states among the lowest
five states in qualitative agreement with data, but not as well
converged.  However, we do not find any low-lying $0^+$ state in our
calculations, in contrast to experiment; we will come back to this
when discussing $^{12}$C below.  Furthermore, our calculations suggest
that there is a $1^+$ state between the second and third $2^+$ excited state.

\begin{figure}[b]
  \center\includegraphics[width=0.9\columnwidth]{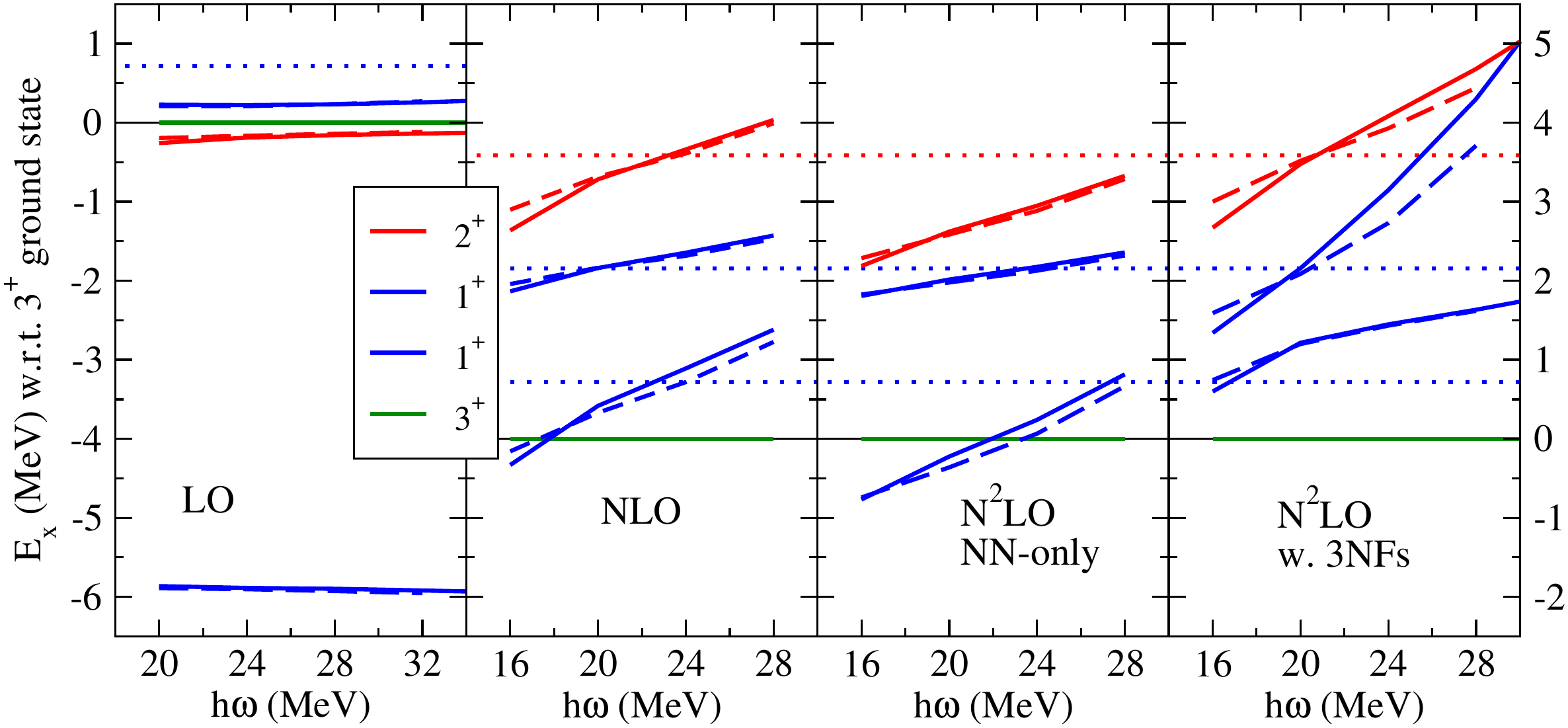}
  \caption{\label{Fig:res_spectrum_10B_hw}
    (Color online) Calculated positive-parity excitation spectrum of
    $^{10}$B using chiral LO, NLO, and N$^2$LO interactions at
    $R=1.0$~fm as function of the basis HO parameter at $\nmax=8$
    for NN-only potentials (left 3 panels) and for N$^2$LO w. 3NFs
    (right-most panel), all SRG-evolved to $\alpha=0.04$~fm$^4$
    (solid) and $\alpha=0.08$~fm$^4$ (dashed).
    Note the different vertical offset for the LO panel.
    The dotted horizontal lines are the experimental values~\cite{TILLEY2004155}.}
\end{figure} 
The low-lying spectra for $^{10}$B up to N$^2$LO are shown in
Fig.~\ref{Fig:res_spectrum_10B_hw}; in addition to the ground state
$3^+$, two low-lying $1^+$ states, and a low-lying $2^+$, there is
also the $0^+$ analog state of the ground state of $^{10}$Be which is
not shown.  At LO the calculated spectrum does not look like the
experimental spectrum at all: the lowest state is a $1^+$ state,
followed by three nearly degenerate states, with $J^P = 2^+$, $3^+$,
and $1^+$, respectively, at excitation energies of about 6 MeV.  At
NLO and NN-only N$^2$LO the agreement with experiment is noticeably
better, except for the ordering of the $3^+$ ground state and the
lowest $1^+$ state.  This is a known issue, and the general consensus
is that 3NFs are needed to achieve the proper $3^+$ ground state for
$^{10}$B~\cite{Navratil:2007we,Jurgenson:2013yya}.  Indeed, adding the
3NFs at N$^2$LO does give the correct ground state, followed by two
$1^+$ states with excitation energies of a few MeV.  However, these
two low-lying $1^+$ states mix, with the amount of mixing strongly
dependent on the basis \hw\ and \nmax\ parameters, which makes it
difficult to extract actual excitation energies for these two
states~\cite{Jurgenson:2013yya}.  The lowest $2^+$ is in reasonable
agreement with the data at N$^2$LO with 3NFs.

\begin{figure}[b]
  \center\includegraphics[width=0.9\columnwidth]{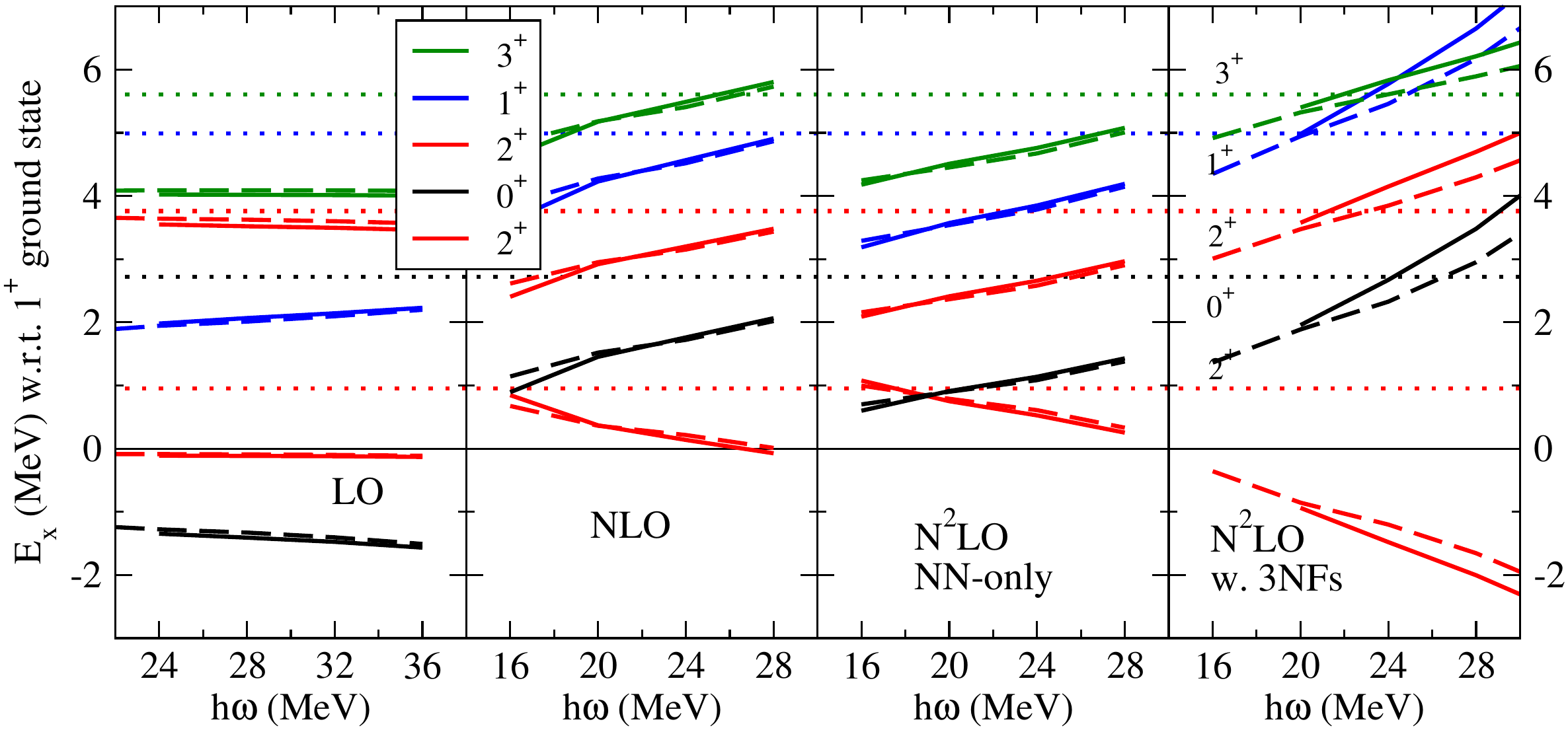}
  \center\includegraphics[width=0.9\columnwidth]{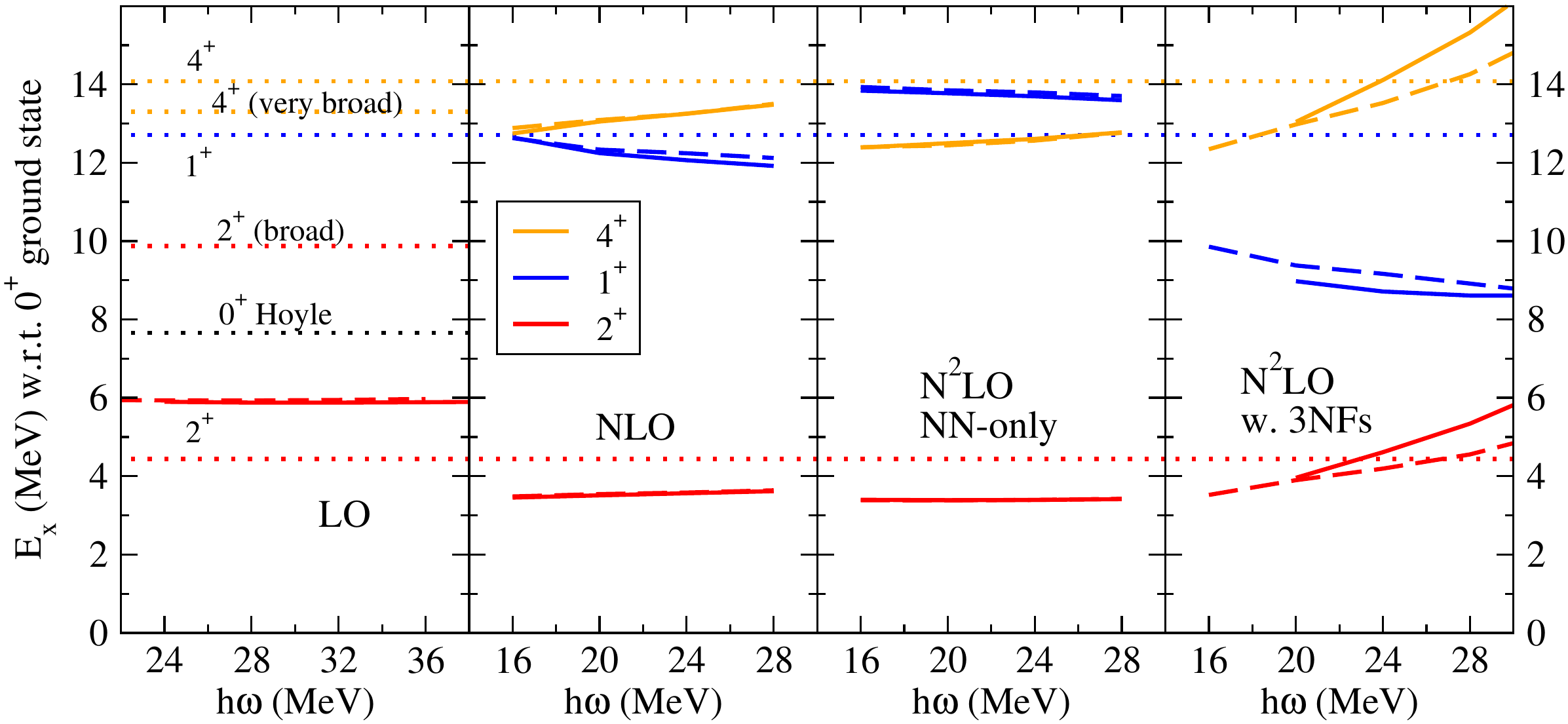}
  \caption{\label{Fig:res_spectrum_12B12C_hw}
    (Color online) Calculated positive-parity excitation spectrum of
    $^{12}$B (top) and $^{12}$C (bottom) using chiral LO, NLO, and
    N$^2$LO interactions at $R=1.0$~fm as function of the basis HO
    parameter at $\nmax=8$ for NN-only potentials (left 3 panels) and
    for N$^2$LO w. 3NFs (right-most panel), all SRG-evolved to
    $\alpha=0.04$~fm$^4$ (solid) and $\alpha=0.08$~fm$^4$ (dashed).
    The dotted horizontal lines are the experimental values~\cite{KELLEY201771}.}
\end{figure} 
Finally, in Fig.~\ref{Fig:res_spectrum_12B12C_hw} we show the
low-lying positive-parity spectra for $^{12}$B and $^{12}$C.  Again,
at LO the spectra do not agree with experiment; furthermore, we do not
find the Hoyle state in $^{12}$C (nor any of its rotational
excitations) due to the known limitations of the HO
basis~\cite{Chernykh:2007zz}.  Furthermore, our spectra at NLO and
N$^2$LO show a significant sensitivity to the chiral order, as well as
the 3NFs at N$^2$LO, for both of these two nuclei.

In particular, at N$^2$LO with 3NFs the first excited $2^+$ state in
$^{12}$B becomes the ground state in our calculations, and the
splitting between this state and the other excited states is
significantly too large.  On the other hand, the energy differences of
the $0^+$, the second $2^+$, the $1^+$, and the $3^+$ relative to the
lowest $1^+$ state are in better agreement with 3NFs than without 3NFs
at N$^2$LO.
Possibly even more puzzling, though not surprising, is the lowest
$1^+$ excited state in $^{12}$C~\cite{Maris:2014hga}.  At NLO it is in
reasonable agreement with experiment, just below the $4^+$ rotational
excitation of the ground state; at N$^2$LO without 3NFs, the order of
the $1^+$ and the $4^+$ is reversed; and including the 3NFs at N$^2$LO
reduces the excitation energy of the $1^+$ by about 5 MeV, destroying
the qualitative agreement with expemiment.  Note that this shift due
to the 3NFs is significantly larger than that for the $2^+$ state in
$^{12}$B, which is of the order of 2 MeV.

In conclusion most spectra for $p$-shell nuclei up to to $A=12$,
calculated at N$^2$LO with 3NFs, agree reasonably well with the
experimental data, in particular for narrow states.  The exceptions
are two states, in $^{12}$B and $^{12}$C respectively.  The low-lying
spectra of $^{10}$B and $^{12}$B, together with the excitation energy
of the lowest $1^+$ state in $^{12}$C, could play a critical role in
determining accurate NN and 3N interactions for the upper $p$-shell
and beyond.  Indeed, both the $2^+$ state in $^{12}$B and the $1^+$
state in $^{12}$C are sensitive to e.g. the LECs $c_D$ and $c_E$.

\ack This work was supported by the US Department of Energy under
Grant No. DE-SC0018223 (SciDAC-4/NUCLEI) and the Funda\c c\~ao de
Amparo \`a Pesquisa do Estado de S\~ao Paulo, Brazil (FAPESP) under
Grant No. 2017/19371-0.  This research used resources of the National
Energy Research Scientific Computing Center (NERSC) and the Argonne
Leadership Computing Facility (ALCF), which are US Department of
Energy Office of Science user facilities, supported under Contracts
No. DE-AC02-05CH11231 and No. DE-AC02-06CH11357, and computing
resources provided under the INCITE award `Nuclear Structure and
Nuclear Reactions' from the US Department of Energy, Office of
Advanced Scientific Computing Research.

\section*{References}

\bibliography{procRTFNB_Maris}{}

\providecommand{\newblock}{}
\begin{thebibliography}{10}
\expandafter\ifx\csname url\endcsname\relax
  \def\url#1{{\tt #1}}\fi
\expandafter\ifx\csname urlprefix\endcsname\relax\def\urlprefix{URL }\fi
\providecommand{\eprint}[2][]{\url{#2}}

\bibitem{Barrett:2013nh}
  Barrett B~R, Navr\'atil P and Vary J~P 2013 {\em Prog. Part. Nucl. Phys.\/} {\bf 69} 131

\bibitem{Maris:2008ax}
  Maris P, Vary J~P and Shirokov A~M 2009 {\em Phys. Rev.\/} {\bf C79} 014308

\bibitem{Coon:2012ab}
  Coon S~A, Avetian M~I, Kruse M~K~G, van Kolck U, Maris P and Vary J~P 2012 {\em Phys. Rev.\/} {\bf C86} 054002

\bibitem{Furnstahl:2012qg}
  Furnstahl R~J, Hagen G and Papenbrock T 2012 {\em Phys. Rev.\/} {\bf C86} 031301

\bibitem{More:2013rma}
  More S~N, Ekstr\"om A, Furnstahl R~J, Hagen G and Papenbrock T 2013 {\em Phys. Rev.\/} {\bf C87} 044326

\bibitem{Wendt:2015nba}
  Wendt K~A, Forss\'en C, Papenbrock T and S\"a\"af D 2015 {\em Phys. Rev.\/} {\bf C91} 061301

\bibitem{doi:10.1002/cpe.3129}
  Aktulga H~M, Yang C, Ng E~G, Maris P and Vary J~P 2014 {\em Concurrency Computat.: Pract. Exper.\/} {\bf 26} 2631

\bibitem{SHAO20181}
  Shao M, Aktulga H, Yang C, Ng E~G, Maris P and Vary J~P 2018 {\em Comput. Phys. Commun.\/} {\bf 222} 1

\bibitem{Weinberg:1990rz}
  Weinberg S 1990 {\em Phys. Lett.\/} {\bf B251} 288

\bibitem{Epelbaum:2008ga}
  Epelbaum E, Hammer H~W and Mei\ss{}ner U~G 2009 {\em Rev. Mod. Phys.\/} {\bf 81} 1773
  
\bibitem{Machleidt:2011zz}
  Machleidt R and Entem D~R 2011 {\em Phys. Rept.\/} {\bf 503} 1

\bibitem{Binder:2015mbz}
  Binder S {\em et~al.\/} (LENPIC) 2016 {\em Phys. Rev.\/} {\bf C93} 044002

\bibitem{Binder:2018pgl}
  Binder S {\em et~al.\/} (LENPIC) 2018 {\em Phys. Rev.\/} {\bf C98} 014002

\bibitem{Epelbaum:2018ogq}
  Epelbaum E {\em et~al.\/} (LENPIC) 2019 {\em Phys. Rev.\/} {\bf C99} 024313
  
\bibitem{Epelbaum:2014efa}
  Epelbaum E, Krebs H and Mei\ss{}ner U~G 2015 {\em Eur. Phys. J.\/} {\bf A51} 53

\bibitem{Epelbaum:2014sza}
  Epelbaum E, Krebs H and Mei\ss{}ner U~G 2015 {\em Phys. Rev. Lett.\/} {\bf 115} 122301

\bibitem{Epelbaum:2002vt}
  Epelbaum E, Nogga A, Gloeckle W, Kamada H, Mei\ss{}ner U~G and Wita\l{}a H 2002 {\em Phys. Rev.\/} {\bf C66} 064001

\bibitem{Nogga:2005hp}
  Nogga A, Navr\'atil P, Barrett B and Vary J~P 2006 {\em Phys. Rev.\/} {\bf C73} 064002

\bibitem{Navratil:2007we}
  Navr\'atil P, Gueorguiev V~G, Vary J~P, Ormand W~E and Nogga A 2007 {\em Phys. Rev. Lett.\/} {\bf 99} 042501

\bibitem{Gazit:2008ma}
  Gazit D, Quaglioni S and Navr\'atil P 2009 {\em Phys. Rev. Lett.\/} {\bf 103} 102502

\bibitem{Bogner:2007rx}
  Bogner S~K, Furnstahl R~J, Maris P, Perry R~J, Schwenk A and Vary J~P 2008 {\em Nucl. Phys.\/} {\bf A801} 21

\bibitem{Bogner:2009bt}
  Bogner S~K, Furnstahl R~J and Schwenk A 2010 {\em Prog. Part. Nucl. Phys.\/} {\bf 65} 94

\bibitem{Roth:2013fqa}
  Roth R, Calci A, Langhammer J and Binder S 2014 {\em Phys. Rev.\/} {\bf C90} 024325

\bibitem{Maris:2013poa}
  Maris P and Vary J~P 2013 {\em Int. J. Mod. Phys.\/} {\bf E22} 1330016

\bibitem{Jurgenson:2013yya}
  Jurgenson E~D, Maris P, Furnstahl R~J, Navr\'atil P, Ormand W~E and Vary J~P 2013 {\em Phys. Rev.\/} {\bf C87} 054312

\bibitem{Audi2003337}
  Audi G, Wapstra A and Thibault C 2003 {\em Nucl. Phys.\/} {\bf A729} 337

\bibitem{Calci:2016dfb}
  Calci A, Navr\'atil P, Roth R, Dohet-Eraly J, Quaglioni S and Hupin G 2016 {\em Phys. Rev. Lett.\/} {\bf 117} 242501

\bibitem{TILLEY20023}
  Tilley D {\em et~al.\/} 2002 {\em Nucl. Phys.\/} {\bf A708} 3

\bibitem{TILLEY2004155}
  Tilley D, Kelley J, Godwin J, Millener D, Purcell J, Sheu C and Weller H 2004 {\em Nucl. Phys.\/} {\bf A745} 155

\bibitem{KELLEY201771}
  Kelley J, Purcell J and Sheu C 2017 {\em Nucl. Phys.\/} {\bf A968} 71

\bibitem{Chernykh:2007zz}
  Chernykh M, Feldmeier H, Neff T, von Neumann-Cosel P and Richter A 2007 {\em Phys. Rev. Lett.\/} {\bf 98} 032501
  
\bibitem{Maris:2014hga}
  Maris P, Vary J~P, Calci A, Langhammer J, Binder S and Roth R 2014 {\em Phys. Rev.\/} {\bf C90} 014314

\end{thebibliography}
\bibliographystyle{iopart-num}

\end{document}